\documentclass[twocolumn,showpacs,floatfix,prl]{revtex4}

\usepackage{float}
\usepackage{graphicx}

\newcommand{\bk}{{\bf k}}
\newcommand{\br}{{\bf r}}

\begin{document}

\title{Topological Crystalline Insulators}

\author{Liang Fu}
\affiliation{Department of Physics, Harvard University, Cambridge, MA 02138}

\begin{abstract}
The recent discovery of topological insulators has revived  interest in the 
band topology of insulators. In this work, we extend the topological classification of 
band structures to include certain crystal point group symmetry. 
We find a class of three-dimensional ``topological crystalline insulators'' 
which have metallic surface states with quadratic band degeneracy on high symmetry crystal surfaces.  
These topological crystalline insulators are the counterpart of topological insulators in  
materials without spin-orbit coupling. Their band structures are characterized by new topological invariants. 
We hope this work will enlarge the family of topological phases in band insulators and stimulate the search for them 
in real materials. 
\end{abstract}

\maketitle

The topology of band structures is important in the study 
of topological phases of matter. Historically the quantization of Hall conductance in the integer quantum 
Hall effect was explained by the Thouless-Kohmoto-Nightingale-den Nijs 
integer (also known as Chern number) of occupied energy bands\cite{thouless}. 
Recently the topological classification of spin-orbital coupled band structures with time reversal symmetry 
has played a key role in the theoretical prediction of topological insulators\cite{kanereview, moorereview, zhangreview}.  
The subsequent development of topological band theory\cite{fukane}, combined with 
realistic band structure calculations, has proven useful in the material search for topological insulators.

Inspired by the discovery of topological insulators,   
the classification of band structures has been extended to other discrete symmetry classes 
such as particle-hole symmetry\cite{ludwig, kitaev, ying}, which leads to a rich family of topological phases such as 
topological superconductors\cite{zhangtsc, ludwig}. More recently, the classification of magnetic insulators with 
certain magnetic translation symmetry has been studied\cite{zhangmagnetic, mooremagnetic}.   Finding 
these phases in real materials is interesting and challenging.  

The purpose of this work is to extend the classification of band structures in a different direction---to 
include crystal point group symmetries. 
We introduce the notion of ``topological crystalline insulators'', which cannot be smoothly 
connected to a trivial atomic insulator when time reversal ($T$) symmetry and certain point group 
symmetry are respected. Unlike time reversal symmetry, crystal symmetries can be broken by sample surfaces. 
As a result, a low-symmetry surface of a topological crystalline insulator does not have robust surface states. 
This motivates us to study a class of three-dimensional topological crystalline 
 insulators which have four-fold ($C_4$) or six-fold ($C_6$) rotational symmetry. As we will show,  
its (001) surface, which preserves the rotational symmetry, supports gapless surface states. 
 These topological crystalline insulators are the counterpart of topological insulators in materials 
{\it without spin-orbit coupling}. Instead electron's orbital degrees of freedom play a role similar to spin.  
Unlike the linearly dispersing Dirac surface states of topological
insulators, the (001) surface states of topological crystalline insulators have {\it quadratic} band degeneracy protected by 
time reversal and discrete rotational symmetry\cite{wen,sun}.

The outline of this paper is as follows. First we introduce a simple
tight-binding model in a tetragonal crystal with $C_4$ symmetry.  
We explicitly show that gapless  
surface states  exist on the (001) crystal face. 
The topological stability of surface states suggests a nontrivial phase 
in this model. 
Next we define a new $Z_2$ topological invariant for generic time reversal invariant 
band structures with $C_4$ or $C_6$ symmetry in three dimensions. This establishes the 
topological crystalline insulator phase. 
 Finally we discuss the electronic properties of quadratic surface bands. 


\begin{figure}
\centering
\includegraphics[width=3in]{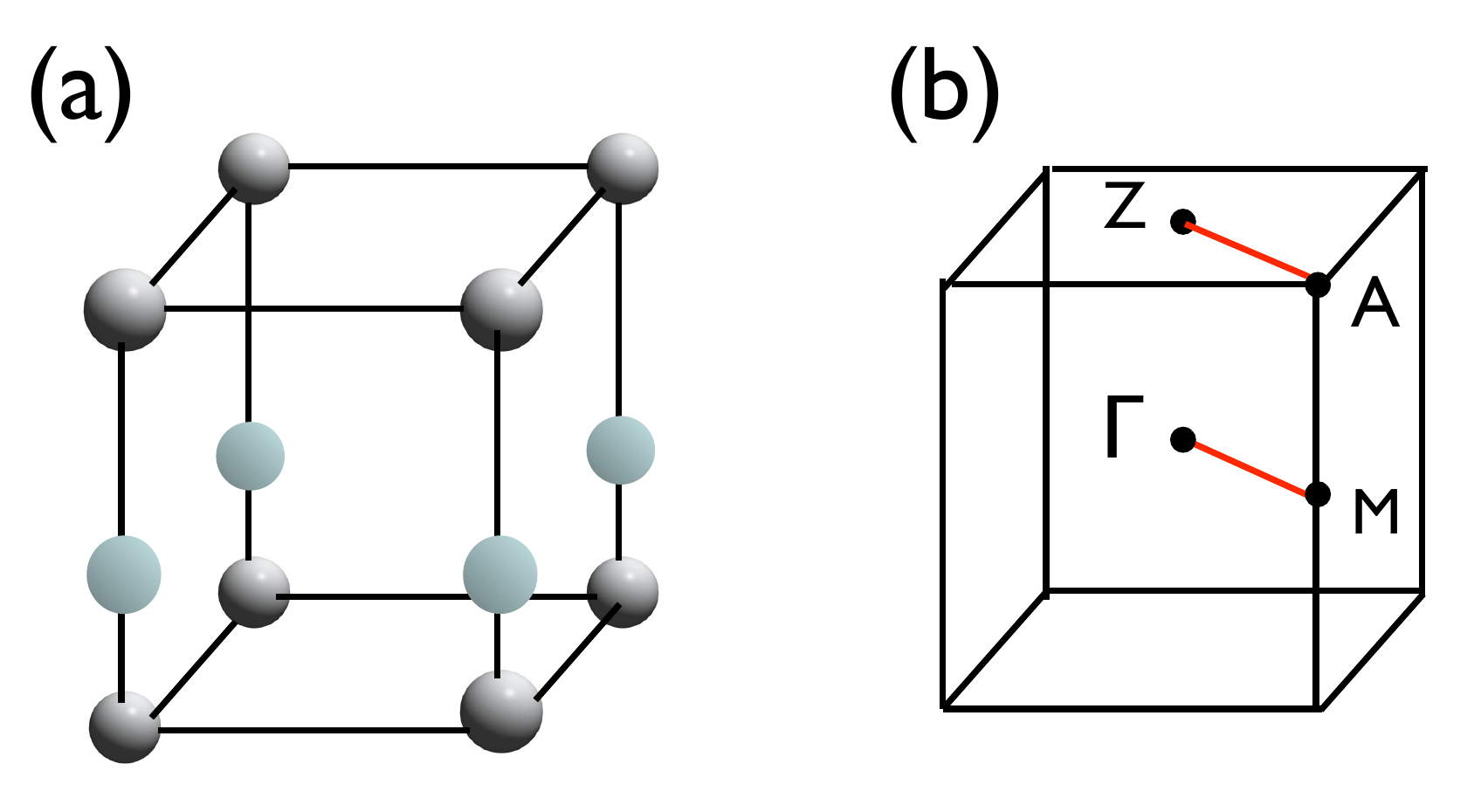}
\caption{(a) tetragonal lattice with two atoms $A$ and $B$ along the $c$-axis in the unit cell; (b) the Brillouin zone and 
four high symmetry points.}
\end{figure}

{\bf Tight-Binding Model:} 
Consider a tetragonal lattice with a unit cell of 
two inequivalent atoms $A$ and $B$ along $c$-axis, as shown in Fig.1a.  
The 3D crystal can be viewed as a stack of bilayer square lattices in the $ab$ plane.  
We now introduce a  tight-binding model to describe the band structure of electron's $p$ orbitals 
(or $d$ orbitals, see the next paragraph).  
In particular, we are interested in the energy bands derived from $p_x$ and $p_y$ orbitals. 
We assume that these bands do not overlap with the $p_z$ bands, 
and construct a tight-binding model from 
Wannier functions with the same symmetry as $p_x$ and $p_y$ orbitals. 
The Hamiltonian $H$ consists of intra-layer hopping $H^A$ and $H^B$, as well as inter-layer hopping $H^{AB}$: 
 \begin{eqnarray}
H &=& \sum_n H^A_n+ H^B_n + H^{AB}_n, \nonumber \\
H^a_n &=&   \sum_{i,j} t^a( \br_i - \br_j ) \sum_{\alpha,\beta} c^\dagger_{a\alpha}(\br_i, n) e^{ij}_{\alpha} e^{ij}_{\beta}  c_{a\beta}(\br_j, n),  \nonumber \\
H^{AB}_n &=& \sum_{i, j} t'( \br_i - \br_j ) [ \sum_{\alpha} c^\dagger_{A\alpha}(\br_i, n)   c_{B\alpha}(\br_j, n) + h.c. ] \nonumber \\
&+& t'_z  \sum_i    \sum_{\alpha} [ c^\dagger_{A\alpha}(\br_i, n)   c_{B\alpha}(\br_i, n+1) + h.c.] \label{H}
\end{eqnarray} 
Here each site is specified by $(n, \br, a)$: 
$n$ labels the bilayer unit cell along $c$-axis; ${\br}=(x, y)$ labels the $ab$-plane coordinate; 
$a=A, B$ labels the sublattice. $\alpha, \beta$ label the $p_x, p_y$ orbital. 
Two types of $p$-orbital hopping terms appear in $H$. The intra-layer hopping in $H^a$ is 
of $\sigma$-bonding type: it depends on the relative orientation 
of $p$-orbital and the hopping direction  ${\bf e}^{ij}=(\br_i - \br_j)/|\br_i - \br_j|$. 
The inter-layer hopping  in $H^{AB}$ is orbital-independent. 
Written in this form, the tight-binding Hamiltonian manifestly preserves crystal symmetries.  
The hopping amplitudes $t^a$, $t'$ and $t'_z$ between two sites 
depend on their $ab$-plane distance $\br_i - \br_j$. Throughout this work, we assume 
spin-orbit coupling is negligible, so that electron's spin index is omitted. 
 
We emphasize that the form of the Hamiltonian $H$ is entirely determined by crystal symmetry. 
Because $d_{xz, yz}$ orbitals transform in the same way as 
$p_{x,y}$ orbitals under $C_4$, (\ref{H}) also applies to materials with these $d$-orbital. Therefore 
(\ref{H}) is potentially revelant to a large class of materials including transition metal compounds with 
$t_{2g}$ orbitals near Fermi energy.

To obtain a minimal model  for topological crystalline insulators, 
we include the nearest and next-nearest neighbor intra-layer hoppings  
in $H^a$ with the amplitude $t^a_1$ and $t^a_2$, as well as  
nearest and next-nearest neighbor inter-layer hoppings in $H^{AB}$ with the amplitude $t'_1$ and $t'_2$. 
The corresponding Bloch Hamiltonian $H(\bf k)$ is obtained by Fourier transform: 
\begin{eqnarray}
H({\bf k}) &=& \left(
\begin{array}{cc}
H^A({\bf k})  & H^{AB}(\bf k) \\
H^{AB \dagger}(\bf k) & H^B({\bf k}) 
\end{array}
\right) \nonumber \\
 H^a({\bf k}) &=& 2 t_1^a \left(
\begin{array}{cc}
 \cos k_x   & 0 \\
0 &  \cos k_y 
\end{array}
\right) \nonumber \\
&+&  2 t_2^a \left(
\begin{array}{cc}
\cos k_x \cos k_y & \sin k_x \sin k_y \\
\sin k_x \sin k_y &  \cos k_x \cos k_y
\end{array}
\right), \nonumber \\
H^{AB}({\bf k}) &=&  [ t'_1   + 2 t'_2 (\cos k_x + \cos k_y) +  t'_z e^{i k_z}] I. \label{Hk}
\end{eqnarray}
The band structure is shown in Fig.2a for the following set of parameters: $t^A_1 = - t^B_1=1 , t^A_2= - t^B_2 = 0.5, t'_1=2.5, t'_2=0.5, t'_z=2$. 
We have checked that the energy gap is finite everywhere in the Brillouin zone. 
As long as the energy gap does not close, the system remains in the same topological class within a finite parameter range.

\begin{figure}
\centering
\includegraphics[width=3in]{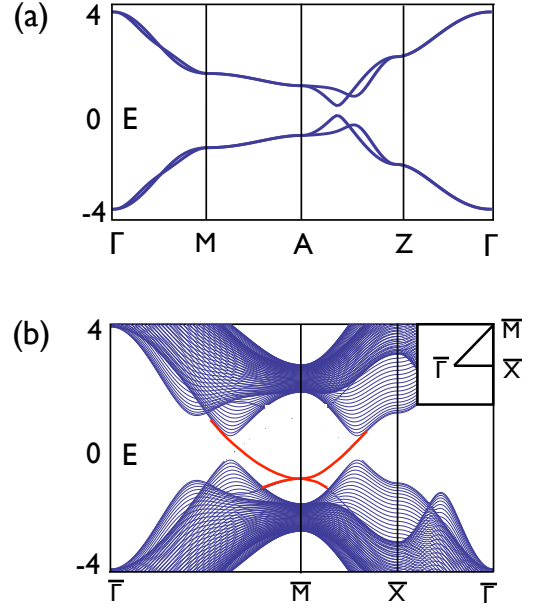}
\caption{(a) bulk band structure of the tight binding model  along high symmetry lines; 
(b) surface states with quadratic band touching exist on (001) face. 
The tight-binding parameters  are shown in the text below Eq.(\ref{Hk}).}
\end{figure}
 
To study surface states, we solve $H$ in a slab geometry. 
We find that the existence of gapless surface states crucially depend on the surface termination. 
Consider the high symmetry (001) surface, which preserves the $C_4$ symmetry. 
Here surface states exist  and transverse the whole energy gap as shown in Fig.2, 
leading to a 2D surface metal.  
Note that surface states are doubly degenerate at $\bar{M}=(\pi, \pi)$: one in $p_x$ orbital 
and the other in $p_y$ orbital.  
Because $\bar M$ is a fixed point under four-fold rotation, 
the doublet form a two-dimensional irreducible {\it real} representation of $C_4$ symmetry, 
as a result of time reversal symmetry for electrons without spin-orbit coupling (i.e., spinless fermions). 
Surface band dispersion near $\bar M$ can be obtained from $k \cdot p$ 
theory. We introduce a pseudospin $\sigma_z=\pm 1$ to label the $p_x$ and $p_y$ orbital of the doublet. 
In this basis, $C_4$ rotation is represented by $e^{i \sigma_y \pi/4}$ and time reversal operator $T$ 
is represented by complex conjugation. 
Up to a gauge transformation, the form of the $k \cdot p$ Hamiltonian $\cal H$ is dictated by symmetry:
\begin{eqnarray}
{\cal H}(k_x, k_y) = \frac{k^2}{2m_0} + \frac{ k_x^2 - k_y^2 }{2m_1} \sigma_z + \frac{k_x k_y}{2m_2}\sigma_x. \label{Heff}
\end{eqnarray}
Perturbations which break either $C_4$ or $T$ symmetry can open up an energy gap 
and destroy the protected surface states. This can be seen explicitly by adding 
a $C_4$-breaking term $M_1 k_x \sigma_y+ M_2 \sigma_z$ or  a $T$-breaking term 
$M \sigma_y$ to  $\cal H$. 
Similar quadratic bands have been recently studied in 2D photonic crystals\cite{wen} and fermion models\cite{sun}.  
Our tight-binding model does not have gapless surface states for other surface terminations which break the $C_4$ symmetry. 

  
When $T$ and $C_4$ symmetry are preserved, the surface states shown above are {\it topologically} protected.   
This can be understood by considering the surface band 
dispersion along $\bar \Gamma \bar M$. 
Within our model, surface states (if present) must be doubly degenerate at $\bar \Gamma$ and $\bar M$. 
As a consequence, 
there exists two distinct  types of surface band connectivity along $\bar \Gamma \bar M$, having
an even or odd number of band crossings at Fermi energy respectively. Surface states with even crossings are fragile: 
they can be pushed out of the energy gap by changing the surface potential. 
Surface states with odd crossings have a Fermi surface which encloses $\bar \Gamma$ an odd number of times. 
They cannot be removed. 
These two types of surface states imply the existence of two topologically distinct phases of time-reversal-invariant band insulators with 
four-fold symmetry. The reasoning here closely follows the previous study of topological insulators\cite{kane}.


{\bf $Z_2$ Topological Invariant}:  
We now show that  a $Z_2$  topological invariant $\nu_0=0, 1$ 
characterizes the band structure of 3D time-reversal-invariant insulators 
with four-fold rotation symmetry (without spin-orbit coupling). 
$\nu_0=0$ corresponds to a trivial phase adiabatically connected to an atomic insulator. 
$\nu_0=1$ corresponds to a topological 
crystalline insulator which has gapless surface states on the $(001)$ surface.    

We start by examining the symmetry property of Bloch wavefunctions of occupied bands: 
$
|\psi_n({\bf k}) \rangle  = e^{ i \bf k \cdot r} | u_n(\bf k) \rangle,
$
where $| u_n (\bf k) \rangle$ is the cell-periodic eigenstates of the Bloch Hamiltonian 
$H({\bf k}) \equiv e^{- i {\bf k \cdot r}} H e^{i {\bf k \cdot r}}$. 
Here the unit cell is chosen to be invariant under $C_4$ rotation around the $z$ axis. 
$H(\bf k)$ satisfies
\begin{eqnarray}
H(k_x, k_y, k_z) &=& U H(k_y, -k_x, k_z) U^{-1} \nonumber \\
H(k_x, k_y, k_z) &=& T H(-k_x, -k_y, -k_z) T^{-1}. \label{T}
\end{eqnarray} 
Here the unitary operator $U = e^{ i \hat{L}_z \pi/2}$ 
implements $C_4$ rotation within the unit cell ($\hat{L}_z=x P_y - y P_x$ is angular momentum operator). 
The anti-unitary operator 
$T=K$ (complex conjugation) implements time reversal transformation for {\it spinless} fermions, 
with the property $T^2=1$. As a result, time reversal symmetry by itself does not 
guarantee a two-fold degeneracy. 
However, the combination of time reversal and four-fold rotational symmetry 
can lead to protected degeneracies at four special momenta  
$\Gamma=(0,0,0), M=(\pi,\pi,0), A=(\pi,\pi,\pi), Z=(0,0,\pi)$ in the 3D Brillouin zone (Fig.1b). 
At such a high symmetry point $\bk_i$,  $H({\bf k}_i)$ commutes with $U$,  
so that the energy bands $|u_n ({\bf k}_i) \rangle$ are eigenstates of four-fold rotation with possible 
eigenvalues $1, -1, i$ and $-i$. Moreover, because $H({\bf k}_i)$ is real, energy bands at $\bk_i$
with $\pm i$ eigenvalues are guaranteed to be degenerate, forming a 
two-dimensional {\it irreducible real} representation of $C_4$.  
For example, such doublet bands can derive from $(p_x, p_y)$ orbitals or $(d_{xz}, d_{yz})$ orbitals. 
From now on, we consider a set of energy bands $| u_n (\bk) \rangle, n=1,...2N$ 
which are doubly degenerate $E_{2n-1}(\bk_i)=E_{2n}(\bk_i)$ at $\Gamma, M, A$ and $Z$. 
As we will see below, only these doublet bands admit a $Z_2$ classification. 
  
The $Z_2$  topological invariant  $\nu$  is defined in terms of the electron wavefunctions: 
\begin{eqnarray}
(-1)^{\nu_0} & = & (-1)^{\nu_{\Gamma M}} (-1)^{\nu_{AZ}}, \\
(-1)^{\nu_{\bk_1\bk_2}} & =& \exp( i \int_{\bk_1}^{\bk_2} d \bk \cdot {\cal A}_\bk) \frac{{\rm Pf} [w(\bk_2)]}{{\rm Pf} [w(\bk_1)]}, \label{it} \\
{\cal A}_\bk &\equiv& -i \sum_j \langle u_j (\bk)| \partial_\bk | u_j (\bk) \rangle, \nonumber \\
 w_{mn}(\bk_i) &\equiv& \langle u_m(\bk_i) | U T | u_n (\bk_i) \rangle \nonumber . 
\end{eqnarray}
${\cal A}_\bk$ is the $U(1)$ Berry connection. $w({\bf k}_i)$ is an antisymmetric U(2N) matrix because
$[H({\bf k}_i), UT]=0$ and $(UT)^2=-1$. $\rm Pf$ stands for the Pfaffian. 
We now specify the integration path in (\ref{it}). For $\nu_{\Gamma M}$, the integral is along an arbitrary line 
connecting $\Gamma$ and $M$ which lies {\it within} the 2D plane $k_z=0$ in the Brillouin zone as shown in Fig.1b.  Likewise, 
the integration path for $\nu_{A Z}$ lies in the plane $k_z=\pi$. 

Both $\cal A$ and $w$ depend on the choice of basis $|u_j(\bk)\rangle$, but we now prove that $\nu_{\bk_1 \bk_2}$ is gauge invariant. 
Two different basis $|\tilde{u}_{n}(\bk) \rangle$ and $|u_{n}(\bk) \rangle$ are related by a $U(2N)$ gauge transformation:
\begin{eqnarray}
|\tilde{u}_{n}(\bk) \rangle =  {\cal G}_{nm}(\bk)  | u_{m}(\bk) \rangle, \; {\cal G} \in U(2N)
\end{eqnarray}
This leads to the corresponding gauge transformation of $\cal A$ and $w$:
\begin{eqnarray}
\tilde{{\cal A}}  &=& {\cal A} - i {\rm Tr} [ {\cal G}^{-1} \partial_\bk {\cal G} ] = {\cal A} - i \partial_\bk \log(\det[{\cal G}] ) \nonumber \\
\tilde{w} &=& {\cal G}^*  w \cal G^\dagger
\end{eqnarray}
Using the identity ${\rm Pf}[X^T M X]=\det[X] {\rm Pf}[M]$, it is straightforward to verify that 
$\tilde{\nu}_{\bk_1 \bk_2} = v_{\bk_1 \bk_2}$ is gauge invariant. 

We now further show that $(-1)^{v_{\bk_1 \bk_2}}=\pm 1$ is a $Z_2$ quantity. 
It follows from (\ref{T}) that at the two 2D planes $k_z=0$ and $k_z=\pi$, $H(\bk)$ has the symmetry property:
$
H(\bk) = \Xi H(\bk) \Xi, \; \Xi \equiv U^2 T. 
$
This allows us to choose a {\it real} gauge along the 
integration path to evaluate (\ref{it}):   
\begin{eqnarray}
\Xi | u_m({\bf k}) \rangle = - | u_m({\bf k}) \rangle,  \label{gauge}
\end{eqnarray} 
Because $\Xi$ is anti-unitary, ${\cal A}=0$ everywhere along the integration path. So in this gauge $(-1)^{\nu[\bk_1, \bk_2]}$ reduces to 
\begin{eqnarray}
(-1)^{\nu_{\bk_1 \bk_2}} = {\rm Pf} [w(\bk_2)]/{\rm Pf} [w(\bk_1)]. 
\end{eqnarray} 
Because $ | u_m(\bk_i) \rangle$ belong to the two-dimensional representation of $C_4$ group, 
we have $U^2 | u_m(\bk_i) \rangle = - |u_m(\bk_i) \rangle$, so that the gauge condition (\ref{gauge}) at $\bk_i$ is equivalent to 
$T|u_m(\bk_i) \rangle =  |u_m(\bk_i) \rangle$, i.e., the wavefunction is real. Under this reality condition, the matrix $w(\bk_i)$ simplifies to 
\begin{eqnarray}
w_{mn}(\bk_i) =  \langle u_m(\bk_i) | U   | u_n (\bk_i) \rangle. 
\end{eqnarray}
Now by choosing a particular kind of real basis 
$|u_{2m}(\bk_i)\rangle \equiv U |u_{2m-1}(\bk_i)\rangle $, 
$w(\bk_i)$ reduces to a standard form $w^0=\epsilon  \oplus \epsilon ...\oplus \epsilon$, 
which is a direct sum of $N$ $2\times 2$ Levi-Civita tensors. This means that in a generic real basis, $w(\bk_i)$ can be written as  
\begin{eqnarray}
w(\bk_i) = O^T(\bk_i) w^0 O(\bk_i), \; O(\bk_i) \in  O(2N).
\end{eqnarray}  
So we have 
\begin{eqnarray}
{\rm Pf} [w(\bk_i)] = \det[O(\bk_i)] {\rm Pf} [w^0] = \det[O(\bk_i)] = \pm 1.  
\end{eqnarray} 
This proves that $(-1)^{\nu_{\bk_1 \bk_2}}=\pm 1$ is a gauge invariant $Z_2$ quantity. 
So both 
$\nu_{\Gamma M}$ and $\nu_{A Z}$ are $Z_2$ topological invariants which characterize the 
band structures in the 2D momentum space $k_z=0$ and $k_z=\pi$ respectively.   
Their product $\nu_0$ is defined for 3D time-reversal-invariant insulators with $C_4$ symmetry, and 
determines the surface state property:  gapless surface states exist on (001) face only when $\nu_0=1$. 
This is verified in the tight-binding model (\ref{H}). 
The relation between $\nu_0$ and $\nu_{\Gamma M}, \nu_{A Z}$ is analogous to that of strong and weak $Z_2$ index 
in 3D topological insulators\cite{fkm, moore, roy}. 

{\bf Generalization to Crystals with $C_6$ Symmetry}: The study of topological crystalline insulators with $C_4$ symmetry 
can be generalized to hexagonal crystal structures with $C_6$ symmetry. 
In that case, (001) surface states have quadratic degeneracy points either 
at $\bar \Gamma$ or two equivalent $\bar K$. 
The topological invariant is similarly defined by Eq.(\ref{it}) 
from the electron wavefunctions on the lines $\Gamma K$ and $AH$ in the Brillouin zone. 
A concrete tight-binding model can be obtained by placing (\ref{H}) in a layered triangular lattice.  
The details will be reported elsewhere. 

The fact that the above $Z_2$ invariant is only defined for doublet bands illustrates the important interplay between 
symmetry representation and topology of energy bands in solids, a feature absent in other known classes of topological insulators. 
We are not aware of 
a systematic mathematical approach to classifying vector bundles (e.g., electron wavefunctions over the Brillouin zone) 
with given representations of group action (e.g., $C_4$ rotation) at fixed points (e.g. high symmetry momenta $\bk_i$).  
We hope the concrete topological invariant (\ref{it}) can  
stimulate continuing research on topological crystalline insulators with other lattice structures. 

The occupied bands of real materials usually have both doubly degenerate and nondegenerate singlet components. 
Provided that the doublet bands are energetically separable from the singlet ones (i.e., no band crossing 
between them), the $Z_2$ invariant remains well-defined for the former. The coexistence with 
singlet bands can in principle weaken the stability of surface states (see supplementary material), 
although this senario seems unlikely to happen in reality. 

{\bf Surface States with Quadratic Degeneracy}: 
Quadratic degeneracies in 2D band structure have attracted considerable interest recently.  
The $k \cdot p$ Hamiltonian (\ref{Heff}) is widely used as a low energy approximation for the band structure  
of bilayer graphene near $\pm K$ points. 
However, unlike topological crystalline insulators, the quadratic degeneracy there is {\it not} protected by symmetry.   
Instead, the band dispersion very close to $\pm K$ becomes linear due to trigonal warping effects. 
  
Electron interactions can drive 2D bands with quadratic degeneracy into a variety of interesting broken symmetry phases including quantum 
anomalous Hall state, nematic phase and etc\cite{sun}. The competition between different ordered phases is of great interest in bilayer graphene. 
It remains to see what happens in the {\it single-valley} quadratic surface states of 
topological crystalline insulators.  

It would be interesting to generalize the concept of topological crystalline insulators to interacting systems. 
For example, crystal symmetry is known to play an important role in the topological classification of spin liquids\cite{spin}, 
spin chains\cite{AKLT}, and Mott insulators\cite{yao}. A unifying theory of symmetry protected 
topological phases remains to be developed. We also note that analogs of topological crystalline insulators
can be potentially realized in photonic crystals.


Part of this work was done in early 2009 at University of Pennsylvania under the support of NSF grant DMR-0906175. 
We thank Charlie Kane and Jeffrey Teo  for very helpful discussions, and acknowledge support from the Harvard Society of Fellows.

\section{Supplementary Material}

We study the stability of surface states with quadratic degeneracy when the doublet bands coexist with 
singlet band in the bulk. The latter may come from, e.g., $p_z$ orbitals so far neglected 
in our tight-binding model (\ref{H}) on tetragonal lattice. Provided there is no band crossing 
between doublet and singlet bands, the $Z_2$ invariant $\nu$  for the former remains well-defined. 
However, as we now demonstrate, the quadratic surface states of a nontrivial $\nu=1$ phase can in principle be 
damaged by hybridizing with additional singlet surface band.    

We assume that the singlet band in the bulk gives rise to a branch of unprotected nondegenerate surface state with 
the usual dispersion $E_3(k)=\epsilon_3 + k^2/(2m_3)$.  $\epsilon_3$ is the energy difference between the 
doublet and singlet surface state at $k=0$. 
This extra surface state can hybridize with   
the original surface states with quadratic degeneracy. A hybridization term is symmetry allowed in the $k \cdot p$ 
Hamiltonian: 
\begin{eqnarray}
{\cal H}' &=& \left( 
\begin{array}{cc}
{\cal H} & \begin{array}{c} 0 \\ 0 \end{array} \\
0 \;\; 0 &  \epsilon_3 +\frac{k^2}{2m_3} 
\end{array} \right) 
+ \lambda
\left(
\begin{array}{ccc}
0 & 0 & i k_x \\
0 & 0 & i k_y \\
- i k_x & -i k_y & 0
\end{array} \right). \nonumber 
\end{eqnarray}
 Here $\cal H$ is the $2 \times 2$ Hamiltonian defined in Eq.(\ref{Heff}). 
 $\lambda$ is the coupling strength between them responsible for hybridization.  
  For $\epsilon_3 m_3 <0$ and $\lambda=0$, singlet and doublet surface bands cross at $k \neq 0$.      
 A finite coupling $\lambda \neq 0$ turns the band crossing into 
 an avoided crossing and opens up an energy gap. The resulting 
 surface states cease to be protected: they can be smoothly pushed into the bulk continuum. 
 This shows that the intervention from singlet bands can in principle destroy the gapless surface 
 state with quadratic degeneracy. 
  However, the above scenario requires the coexistence of singlet surface state under particular 
  conditions, which seems very nongeneric in real materials. Therefore we are optimistic that 
  surface states with quadratic degeneracy on topological crystalline insulators are robust for practical 
  purposes.


\begin{thebibliography}{10}

\bibitem{thouless}
D. J. Thouless {\it et al.}, Phys. Rev. Lett. {\bf 49}, 405 (1982).




\bibitem{kanereview} 
M. Z. Hasan and C. L. Kane, arXiv:1002.3895

\bibitem{moorereview} 
J. E. Moore, Nature {\bf 464}, 194 (2010).

\bibitem{zhangreview} 
X. L. Qi and S. C. Zhang, arXiv:1008.2026

\bibitem{fukane} 
L. Fu and C.L. Kane, Phys. Rev. B {\bf 76}, 045302 (2007).


 \bibitem{ludwig}
 A. Schynder, S. Ryu, A. Furusaki and A. Ludwig, Phys. Rev. B \textbf{78}, 195125 (2008).

\bibitem{kitaev} A. Kitaev, arXiv:0901.2686 

\bibitem{ying} Y. Ran,  arXiv:1006.5454 


\bibitem{zhangtsc} X. L. Qi, {\it et al}, Phys. Rev.
Lett. \textbf{102}, 187001 (2009); 
M. M. Salomaa and G. E. Volovik, Phys. Rev. B 37, 9298 (1988); R. Roy, arXiv:0803.2868  

\bibitem{mooremagnetic} 
R. K. Mong, A. M. Essin and J. E. Moore, Phys. Rev. B \textbf{81}, 245209 (2010).


\bibitem{zhangmagnetic}
See also, e.g., R. Li {\it et al}, Nat. Phys. {\bf 6}, 284 (2010).



\bibitem{wen}
Y. D. Chong, X.-G. Wen, and M. Soljacic, Phys. Rev. B, {\bf 77}, 235125 (2008). 

\bibitem{sun}
K. Sun {\it et al}, Phys. Rev. Lett. {\bf 103}, 046811 (2009). 




\bibitem{kane}
C. L. Kane, Nat. Phys. {\bf 4}, 348 (2008).

\bibitem{fkm}
L. Fu, C. L. Kane and E. J. Mele,  Phys. Rev. Lett. {\bf 98}, 106803 (2007). 


\bibitem{moore} 
J.E. Moore and L. Balents, Phys. Rev. B {\bf 75}, 121306(R) (2007).

\bibitem{roy} 
R. Roy, Phys. Rev. B. {\bf 79}, 195322 (2009)






\bibitem{spin}
X. -G. Wen, Phys. Rev. B {\bf 65}, 165113 (2002). 

\bibitem{AKLT}
Z. Gu and X. G. Wen, Phys. Rev. B {\bf 80}, 155131 (2009); 
F. Pollman {\it et al}, Phys. Rev. B {\bf 81}, 064439 (2010). 

\bibitem{yao}
H. Yao and S. A. Kivelson, Phys. Rev. Lett. {\bf 105}, 166402 (2010). 


\end{thebibliography}
\end{document}